\documentclass[amsmath, amsfonts, showpacs, twocolumn,prb]{revtex4}
\usepackage{graphicx,color}
\usepackage{epsfig}
\usepackage{bm}
\usepackage{dcolumn}
\usepackage{amsmath}
\usepackage{amssymb}
\usepackage{euscript}
\usepackage{dsfont}

 \newcommand{\sgn}{\operatorname{sgn}}
 \newcommand{\erf}{\operatorname{erf}}
 \newcommand{\AiryAi}{\operatorname{Ai}}

\begin{document}

\title{Interaction-induced backscattering 
 in short quantum wires}

\author{M.-T.~Rieder$^1$,  T.~Micklitz$^{1,2}$, A.~Levchenko$^{3,4}$ and
K.~A.~Matveev$^5$}
\affiliation{$^1$Dahlem Center for Complex Quantum Systems and Institut 
f\"ur Theoretische Physik, Freie Universit\"at Berlin, 14195 Berlin, Germany\\
$^2$Centro Brasileiro de Pesquisas F\'isicas, Rua Xavier Sigaud 150, 22290-180, Rio de Janeiro, Brazil\\ 
$^3$Department of Physics and Astronomy, Michigan State University, East
Lansing, MI 48824, USA\\
$^4$Institut f\"{u}r Nanotechnologie, Karlsruhe Institute of Technology, 76021 Karlsruhe, Germany\\
$^5$Materials Science Division, Argonne National Laboratory, 
Argonne, Illinois 60439, USA}

\begin{abstract}
  We study interaction-induced backscattering in clean quantum wires
  with adiabatic contacts exposed to a voltage bias. Particle 
  backscattering relaxes such systems to a fully equilibrated steady
  state only on length scales exponentially large in the ratio of
  bandwidth of excitations and temperature.  Here we focus on shorter
  wires in which full equilibration is not accomplished. Signatures of
  relaxation then are due to backscattering of hole excitations close
  to the band bottom which perform a diffusive motion in
  momentum space while scattering from excitations at the Fermi level.
  This is reminiscent to the first passage problem of a Brownian
  particle and, regardless of the interaction strength, can be
  described by an inhomogeneous Fokker-Planck equation.  From general
  solutions of the latter we calculate the hole backscattering rate
  for different wire lengths and discuss the resulting length
  dependence of interaction-induced correction to the conductance of
a clean single channel quantum wires.
\end{abstract}

\date{\today}


\maketitle

\section{Introduction}

The study of equilibration in many-particle quantum systems has moved
into the focus of recent research
interest.~\cite{Polkovnikov2011,Imambekov2012} This interest has been
partially driven by the impressive experimental progress in realizing
and manipulating many-particle quantum systems.  One remarkable
example is the recent cold atom realization of the Tonks Girardeau
gas, which allowed to study the suppression of relaxation in an
integrable many-body system.~\cite{weiss}
 
Clean mesoscopic quantum wires provide another example of systems in
which equilibration is strongly suppressed.~\cite{Barak2010} 
Specifically, in clean single channel quantum
wires equilibration is due to backscattering of excitations which
occur at energies of the order of their bandwidth
$\Delta$.~\cite{feq,peq,eqLL,GLL,eqWC} As a consequence, the
equilibration rate displays activated temperature dependence, 
$\tau_{\rm eq}^{-1}\propto e^{-\Delta/T}$ and, when exposed
to a finite voltage bias, a fully equilibrated steady state only
occurs in wires exceeding an {\it exponentially large} length scale
$\ell_{\rm eq}\propto e^{\Delta/T}$.

In shorter wires, $L\ll \ell_{\rm eq}$, effects of equilibration on
electrons at the Fermi level can be neglected and signatures of
relaxation are due to backscattering of particles close to the band
bottom.~\cite{Lunde2007,AZT-PRB11} The key process for relaxation in this case
is one in which a thermally activated hole overcomes, as it scatters
from excitations at the Fermi level, a barrier of energetically
unfavorable states at the band bottom via random small steps in
momentum space.  This picture of a Brownian particle applies
regardless of the interaction strength and thus opens the possibility
to study equilibration of one-dimensional fermions beyond the weak
interaction regime. 

Strongly interacting electrons are commonly described within the 
Luttinger liquid framework and previous work 
has studied scattering of a Brownian particle in a homogeneous 
Luttinger liquid.~\cite{CastroFisher}
The focus of the present paper is on the equilibration in voltage biased quantum
wires. Here, the nature of the specific boundary conditions
requires to address a space-dependent, i.e. {\it  inhomogeneous} problem.

The outline of the paper is as follows. Section \ref{sec:backscatt}
introduces the backscattering rate of holes and reviews how a finite
rate affects the conductance of the wire.  In Section~\ref{sec:fokpla}
we discuss the relevant kinetic equation. 
Solutions of the latter, the resulting backscattering
rate and interaction-induced correction to the conductance of the wire 
are discussed in Sections~\ref{sec:idr} and~\ref{sec:longwires}.
Details of the calculations are relegated to the appendices.

\section{Backscattering rate} 
\label{sec:backscatt} 

\begin{figure}[t!]
\centering
\resizebox{.44\textwidth}{!}{\includegraphics{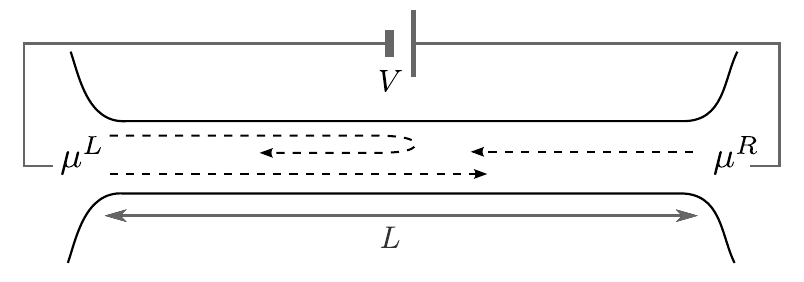}}
 \caption{
 One-dimensional quantum wire of length $L$, adiabatically connected 
to two-dimensional reservoirs which are kept at different equilibria
characterized by 
chemical potentials $\mu^{L/R}=\mu\pm eV/2$. 
Right- and left-moving electrons enter the wire from different 
reservoirs and relax  towards a new 
equilibrium state through interaction-induced backscattering of particles.} 
\label{fig1}
\vspace{-.2cm}
\end{figure}

Consider interacting electrons in a clean one-dimensional quantum wire
adiabatically connected to two-dimensional reservoirs via fully
transparent contacts.  A non-equilibrium situation arises when
reservoirs at left and right contacts are biased by a finite voltage
$V$. Then, right- and left-moving electrons injected into the wire
from the left and right reservoir, respectively, are at different
equilibria, see Fig.~\ref{fig1}.  In the absence of interactions the
conductance of the wire reads
\begin{align}
\label{gnon}
 G_0 = G_q \left(1 - e^{-\mu/T} \right),
\end{align}
where $G_q = 2e^2/h$ is the quantum of conductance, and $\mu$ is the
chemical potential. Accounting for interactions within the Luttinger
liquid framework, the conductance of a finite wire remains
$G_q$.~\cite{maslov,Ponomarenko,Safi} While electrons inside a realistic voltage biased
wire relax towards a new steady state, excitations within the
Luttinger liquid model have an infinite life time. To study effects of
equilibration on finite temperature transport coefficients, one thus
has to go beyond the Luttinger liquid model. As we discuss below, this
can be accomplished even if interactions are strong. 
 
\begin{figure}[t!]
\centering
\resizebox{.46\textwidth}{!}{\includegraphics{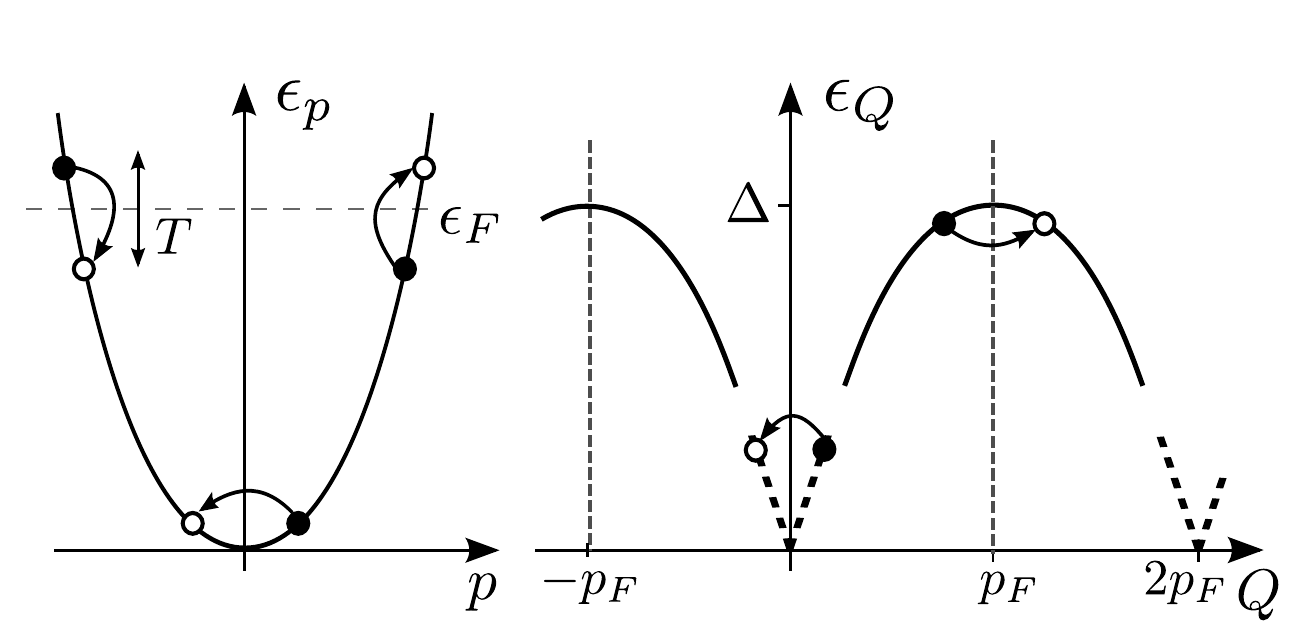}}
\caption{Left: Spectrum of weakly interacting electrons.
Relaxation at weak interactions occurs via
  three-particle scattering processes in the course of which a highly
  excited hole at the band bottom is backscattered. Right: The concept
  of backscattering of a highly excited hole also applies at strong
  interactions.  The energy of a hole excitation $\epsilon_Q$ is shown
  as a function of its momentum $Q$.  First Brillouin zone $|Q|<p_F$
  corresponds to the momentum $Q$ of the hole measured from the nearest
  Fermi point.
  As a result of many collisions with low energy
  bosonic excitations the latter may increase its momentum $Q$ and
  enter the second Brillouin zone, after which it is more likely to
  fall toward $Q=2p_F$ than to return to the vicinity of $Q=0$.  Each
  backscattering event thus corresponds to an umklapp process in which
  a highly excited hole crosses the edge of Brillouin zone (dashed
  lines).  }
\label{fig2}
\end{figure}

Taking into account relaxation into a new steady state,
the latter can be characterized in terms of a local backscattering rate $\dot{n}^R(x)$ of fermions. 
In the limit of weak interactions, $n^R(x)={2\over L}\sum_{p>0}f_{p,x}$ is the density 
of right-moving electrons. The dot here and in the following refers to
the total time derivative,
$f_{p,x}$ is the Fermi distribution of (weakly interacting) electrons
and the factor $2$ is due to spin-degeneracy.  The total
backscattering rate of electrons is then $\dot{N}^R
=\int_{-L/2}^{L/2}dx\, \dot{n}^R(x)$ and a finite rate results from 
backscattering of highly excited holes close to the band bottom, see
Fig.~\ref{fig2}.

The above picture readily generalizes to strong interactions.  Indeed,
one can extend the concept of a hole excitation to arbitrary
interaction strengths by noting that for a system with concave
spectrum the lowest energy excitation at a given momentum $Q$ is a
hole.~\cite{Imambekov2012, Matveev2013} Even though interactions
renormalize its spectrum $\epsilon_Q$, the hole remains a spin-$1/2$
excitation with a two-fold degeneracy of the energy levels protected
by spin-rotation symmetry. The energy spectrum is periodic,
$\epsilon_Q=\epsilon_{Q+2p_F}$, and quasi-momenta of hole excitations
may thus be restricted to the first Brillouin zone $|Q|<p_F$.
Building on this observation, a backscattering event corresponds to an
umklapp process in which the highly excited hole crosses the edge of
the Brillouin zone.  Notice here that the edge of the Brillouin zone
corresponds to the band bottom of the spectrum of the non-interacting
fermions, and both terms will be used synonymously below, see also
Fig.~\ref{fig2}.  In particular, in both pictures the relaxation is
due to processes taking place at the bottom of the band, i.e. 
involving highly excited holes.  To simplify notation, in the
subsequent discussion instead of the momentum of the hole $Q$ measured
from the nearest Fermi point we will use the momentum of the missing
electron near the bottom of the band, $p=p_F{\rm sgn}(Q)-Q$.

Regardless of the interaction strength the total backscattering rate
of holes is calculated in terms of the hole distribution $g_{p,x}$,
 \begin{align}
 \label{bsr}
\dot{N}^R
&=-2\int_{-L/2}^{L/2}dx\,  \sum_{p>0}  \dot{g}_{p,x}.
\end{align}
In the short wires considered in this paper the backscattering rate
(\ref{bsr}) is controlled by momenta close to the band bottom $|p|\ll
p_F$.  We will discuss the hole occupation numbers $g_{p,x}$ in the
following sections.  The factor of 2 in Eq.~\eqref{bsr} again results
from spin-degeneracy, and the minus sign from expressing the
backscattering rate in terms of the hole distribution.  A finite
backscattering rate manifests in a reduced steady state
current,\cite{peq,GLL}
\begin{align}
I=G_0 V+e\dot{N}^R,
\end{align} 
corresponding to a conductance which differs from
that of a non-interacting system \eqref{gnon}, 
\begin{align}
\label{gdg}
G=G_0+\delta G, \quad
\delta G= {e\dot{N}^R\over V}.
\end{align}

The backscattering rate \eqref{bsr} recently has been studied in the
limit of relatively short~\cite{Lunde2007,AZT-PRB11} ($L\ll \ell_0$) and
relatively long ($L\gg \ell_1$) wires.\cite{peq}  While the characteristic
scales $\ell_0$ and $\ell_1$ are discussed in detail below, we mention
here that the condition $L\lesssim \ell_0$ defines a quasi-ballistic
regime, in which a hole at the bottom of the band typically scatters
at most once from excitations at the Fermi level during its passage
through the wire.  On the other hand, the condition $L\gg \ell_1$
defines a homogeneous diffusive regime where the hole suffers from
many collisions granting fully diffusive dynamics in momentum space on
a scale set by temperature.  In both cases the distribution function
of the holes at the bottom of the band is space-independent. That is,
one effectively deals with a homogeneous problem and consequently
linear dependences $\dot{N}^R\propto L$ are found. Slopes, however,
are parametrically different in the quasi-ballistic and homogenous
diffusive regimes. One should therefore expect a nontrivial length
dependence $\delta G(L)$ in the intermediate regime 
which at low temperatures of interest defines a wide region of length scales   
$\ell_0\ll L\ll \ell_1$ specified below. 
As we discuss next, this condition defines an {\it inhomogeneous}
diffusive regime, in which the typical range of diffusive dynamics in
momentum space is set by the length of the wire.  This latter is
addressed within an inhomogeneous Fokker-Planck equation.

\section{Fokker-Planck equation} 
\label{sec:fokpla}

\begin{figure}[tb]
\centering
\resizebox{.46\textwidth}{!}{\includegraphics{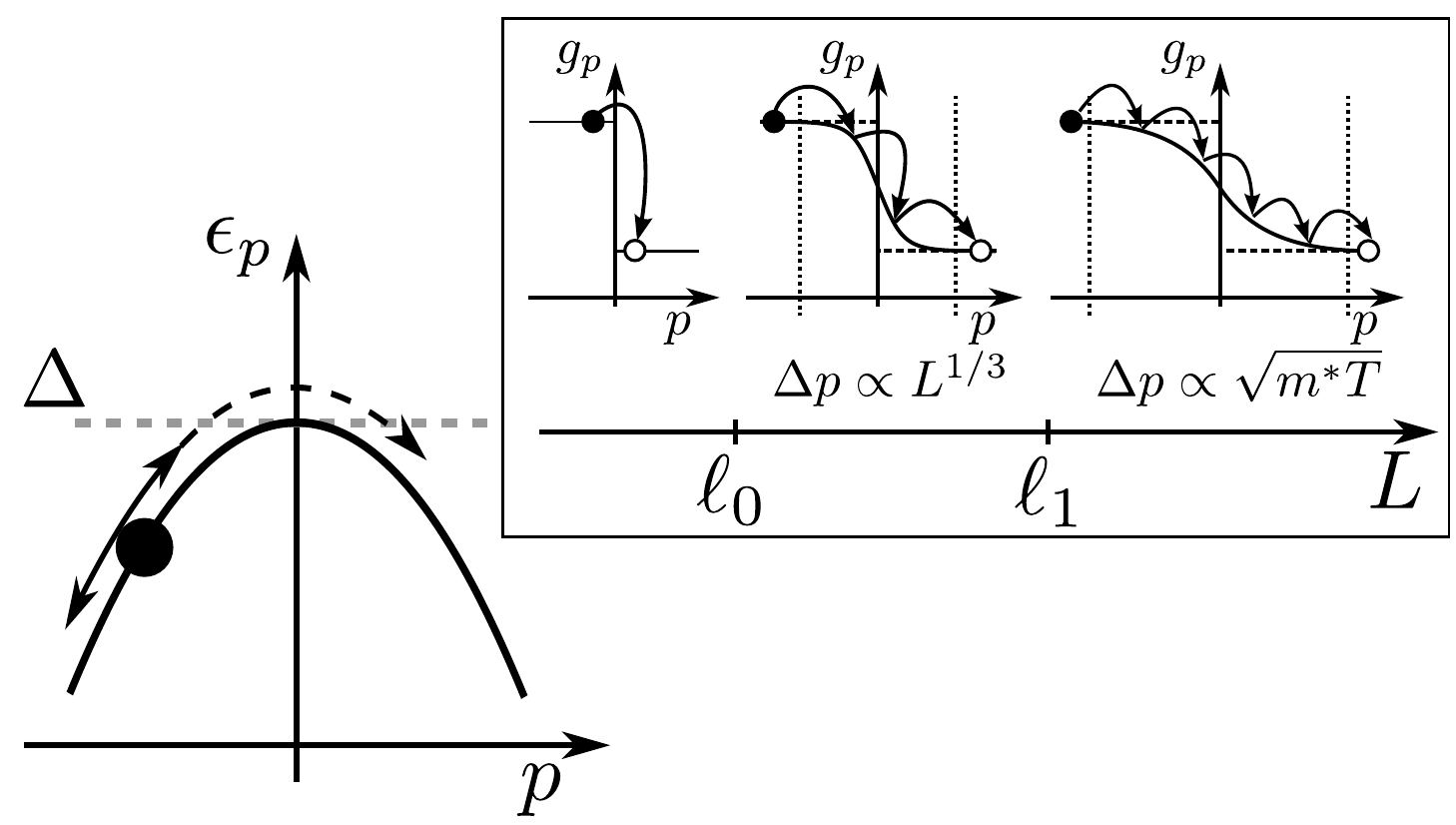}}
\caption{
A hole in the vicinity of the bottom of the band with dispersion Eq.~\eqref{disp} 
performs a random walk in momentum space. 
In the process of equilibration the holes reverses its direction. The backscattering occurs 
in a sequence of collisions with excitations at the Fermi level,
in which a small relative momentum $\delta p/p_0\ll 1$ is transferred to the hole. 
Inset: Evolution of the hole-distribution at small momenta    
as the wire length is increased. In the quasi-ballistic regime of relatively short wires $L\lesssim \ell_0$ a 
hole close to the  band bottom 
suffers on average less than one collision when traversing the wire. 
The voltage induced jump at $p=0$ in this regime is 
 not affected by backscattering and remains 
of the order $\sim e^{-\Delta/ T} eV /T$. Once   
$L  \gg \ell_0$ such hole typically experiences many
collisions during its passage through the wire which
turn its motion in momentum space diffusive. 
For wires in the inhomogeneous diffusive regime, $\ell_0\ll L\ll \ell_1$, 
holes then redistribute  
in a momentum range scaling with the length of the wire as   
$\Delta p \propto L^{1/3}$. Only upon
entering the homogeneous diffusive regime, $L \gg \ell_1$,
this range saturates at a momentum-scale set by temperature
$\Delta p \sim \sqrt{m^*T}$, implying that
the hole-distribution becomes a smooth function 
on this scale.
}
\label{fig3}
\end{figure}

The Fokker-Planck equation (FPE) describes the paradigmatic situation
in which a heavy `Brownian' particle propagates in a dilute gas of
light particles.  Collisions between heavy and light particles then
lead to a diffusive motion of the former.  In our context, the typical
momentum $\delta p$ transferred in a collision between a hole at the
band bottom and thermally excited electron-hole excitations (plasmons)
is restricted due to Fermi blocking to be of the order of $\delta p
\sim T/v$, where $v$ is the velocity of excitations at the Fermi level.
Below we consider the case in which the dispersion of a hole at the
bottom of the band is quadratic,
\begin{align}
\label{disp}
\epsilon_p=\Delta - {p^2\over 2m^*},
\end{align} 
with $m^*$ being the effective mass of the hole.  We then encounter
the above situation at low enough temperatures where the typical
momentum of the hole $p_0\sim\sqrt{m^*T}\gg \delta p$, see also
Fig.~\ref{fig3}.

Formally, we start out from the Boltzmann equation
 \begin{align}
 {p\over m^*}\partial_x g_{p,x}=I_{p,x}[g]
 \end{align} 
 for the hole distribution $g_{p,x}=1-f_{p,x}$ in a steady state
 $\partial_t g_{p,x}=0$. 
 Employing the small parameter $\delta p/p_0\ll 1$ one may perform a
 Kramers-Moyal expansion and approximate the collision integral by the
 Fokker-Planck form
\begin{align}
\label{fpo}
I_{p,x}[g] \simeq  
- \partial_p \left( 
A(p) g_{p,x} 
-  {1\over 2} \partial_p  [B(p) g_{p,x}] 
\right).
\end{align}
The Fokker-Planck operator describes an interplay of drift and
diffusion which sends the system into the new steady
state. Coefficients $A(p)$ and $B(p)$ in Eq.~\eqref{fpo} are model
specific functions.  In all cases of interest, variation of the
coefficient $B(p)$ occurs on a momentum scale much larger than $p_0$,
and it may thus be approximated by a constant $B(p)=B$.  Employing
common statistical mechanics arguments, we further know that in a
homogeneous equilibrium situation, $eV=0$, the dilute hole at the band
bottom is described by a Boltzmann distribution.  This fixes
$A(p)=Bp/(2m^*T)$, and the (dilute) hole distribution thus follows the
space-dependent FPE, also known as Kramers
equation~\cite{kramers,risken}
\begin{align} 
\label{fpeq}
{p\over m^*} \partial_x g_{p,x} 
&= \frac{B}{2} \partial_p 
\left( - {p g_{p,x} \over m^* T} + \partial_p g_{p,x} \right).
\end{align} 
All microscopic details are stored in the single constant $B$, which physically 
speaking has the meaning of a diffusion constant in momentum space. It can be explicitly calculated in the special cases of either weak or strong interactions.~\cite{peq,eqWC,AZT-PRB11,MatveevAndreevKlironomos} Interestingly, one can also obtain a phenomenological expression for $B$ in terms of the spectrum of the mobile impurity (hole) in the Luttinger liquid.~\cite{eqLL,Matveev2013,Matveev2012}
Inhomogeneity in Eq.~\eqref{fpeq} is induced by the boundary conditions taking into account 
the finite voltage bias. For a dilute hole at the band bottom the latter 
can be approximated by a classical Boltzmann form 
\begin{subequations}
 \label{bc}
 \begin{eqnarray}
   \label{eq:bc_left_lead}
     g_{p,-L/2} &=& e^{\epsilon_p/T}e^{-(\Delta+eV/2)/T},\quad\mbox{for $p>0$},\\
   \label{eq:bc_right_lead}
     g_{p,L/2} &=& e^{\epsilon_p/T}e^{-(\Delta-eV/2)/T},\quad\mbox{for $p<0$},
 \end{eqnarray}
\end{subequations} 
where $\Delta$ is the bandwidth of the hole excitations, see Eq.~\ref{disp}.

While the above expressions \eqref{eq:bc_left_lead} and
\eqref{eq:bc_right_lead} are obvious in the limit of weak
interactions, let us notice that independent of the interaction
strength the occupation of (dilute) high-energy excitations in a fluid
at rest is given by the Boltzmann factor $g_{Q}=e^{-\epsilon_Q/T}$,
where $\epsilon_Q$ the excitation spectrum.  A finite voltage bias
sets the fluid in motion and changes the excitation spectrum of the
Galilean invariant system in the stationary frame according to
$\epsilon_Q\to \epsilon_Q+ uQ$ where $u=I/(en)$ is the fluid velocity
expressed in terms of the electric current $I$ and the particle 
density $n$. Restricting then $|Q|\leq p_F$ to the first Brillouin zone
and measuring momenta from the zone boundary we substitute
$Q=p_F\sgn(p)-p$, where for our
purposes $|p|\sim\sqrt{m^*T} \ll p_F$.  Substituting further particle
density $n=4p_F/h$ and current $I\simeq G_qV$, one arrives at the
above boundary conditions.

Using Eq.~\eqref{fpo} one finds the backscattering rate in the
Fokker-Planck approximation
\begin{align}
\label{bsFP}
\dot{N}^R={B\over h}\int_{-L/2}^{L/2}dx\, \left(\partial_pg_{p,x}\right)|_{p=0}.
\end{align} 
Here $\left(\partial_pg_{p,x}\right)|_{p=0}$ affords the
interpretation of the current of the holes in momentum space through
the band bottom, resulting in the interaction-induced correction
$\delta G(L)$ to the conductance of non-interacting electrons, see
Eq.~\eqref{gdg}.

Before we start a detailed analysis of the Kramers equation
\eqref{fpeq}, it is instructive to obtain the characteristic scales of
distance $\ell_1$ and momentum $p_0$ inherent to it.  Assuming that
the expression in the left-hand side of Eq.~(\ref{fpeq}) is of the
same order of magnitude as each of the terms in the right-hand side,
we obtain 
\begin{align}
   \frac{p_0}{m^*\ell_1}
   \sim\frac{B}{m^*T}
   \sim\frac{B}{p_0^2}.
\end{align}
The above conditions are satisfied for $p_0\sim \sqrt{m^* T}$ and
$\ell_1\sim \sqrt{m^*T^3}/B$.  The two scales can be understood as
follows.  The boundary conditions (\ref{bc}) for the hole distribution
function are discontinuous at $p=0$.  In the presence of scattering in
the wire, $B\neq0$, the discontinuity smears.  Such smearing is weak
in short wires, such that $L\ll \ell_1$.  In this case the typical
momentum scale of the smeared distribution is small compared to $p_0$,
and grows with the length of the wire.  In wires longer than $\ell_1$
the smearing reaches its final value $p_0$ dictated by the
temperature of the system and mass of the holes, but not the
scattering rate, see also Fig.~\ref{fig3}.  We start the detailed
analysis of the Kramers equation \eqref{fpeq} with boundary conditions
\eqref{bc} with the study of the inhomogeneous diffusive regime $L\ll \ell_1$.

\section{Inhomogeneous diffusive regime} 
\label{sec:idr}

To quantify the above qualitative considerations let us return to Kramers equation 
\eqref{fpeq} and restrict to short wires $\ell_0\ll L\ll \ell_1$ in the diffusive regime 
and small momenta, specified momentarily. 
We then observe that in this inhomogeneous diffusive regime
the Fokker-Planck operator is dominated by the second derivative
`smearing'-operator, and thus a simplified
analysis applies where drift is neglected, 
\begin{equation}
 \label{Keq}
\frac{p}{m^*} \partial_x g_{p,x}   
=\frac B2 \partial^2_p g_{p,x} \, .
\end{equation}
Indeed, if the discontinuity of the hole distribution occurs on a  scale 
$\Delta p\ll \sqrt{m^*T}$ in momentum space 
one may estimate $\partial_p g\sim g/\Delta p$. Neglecting drift then 
amounts to dropping contributions  $p \Delta p /m^*T\ll 1$ much smaller than
unity 
for small momenta $p\sim \Delta p$ of interest.~\cite{remark2}

It is convenient to define the length scale $\ell_1=\sqrt{8m^*T^3}/ B$
and to introduce the dimensionless variables,
\begin{align}
\label{dimlessv}
q={p\over \sqrt{2m^*T}},
\qquad 
y={x\over \ell_1},
\qquad
\Lambda={L\over 2\ell_1},
\end{align}  
in terms of which Eq.~\eqref{Keq} takes the form
\begin{align}
\label{eq:dimensionless_form}
 \left( \partial_q ^{2} - 2 q \partial_y \right) g_{q,y} = 0.
\end{align}
Let us now insert the separation ansatz
\begin{align}
\label{anskeq}
g_{q,y} = \int_{-\infty}^{\infty} da\, b(a) e^{ay} \varphi_{a,q} \, ,
\end{align} 
where the functions
$\varphi_{a,q}$ satisfy the differential equation
\begin{align}
 \left( \partial_q ^{2} - 2 q a \right) \varphi_a(q) = 0.
\end{align}
Then solutions of Eq.~(\ref{eq:dimensionless_form})
in the linear response regime assume the form
\begin{align}
\label{gairy}
 g_{q,y} = \int_{0}^{\infty} da\, b(a)\left( 
 e^{ay} \varphi_a(q)  - e^{-ay} \varphi_a(-q) 
 \right) \, ,
\end{align} where 
\begin{align}
\varphi_a(q)=(2a)^{-2/3}{\rm Ai}((2a)^{1/3}q)
\end{align} 
and ${\rm Ai}(x)$ the Airy function.  In obtaining
Eq.~\eqref{gairy} we took advantage of the fact that in the linear
response regime the distribution in the center of the wire $g_{q,0}$
is antisymmetric in $q$.  Finally, coefficients $b(a)$ are fixed by
imposing 
the boundary conditions,
\begin{align}
\label{bc2}
 \theta (q) g_{q,-\Lambda} + \theta (-q) g_{q,\Lambda} = -e^{-\Delta/T}
\frac{eV}{2T} \sgn (q). 
\end{align}  
For a detailed discussion on this procedure we refer to
Appendix~\ref{app:ballistic} and move on to the physical implications
of our solution.

Our result for the  backscattering rate at $\ell_0\ll L\ll \ell_1$ has the form
\begin{align}
\label{swl}
\dot{N}^R
&=
- \zeta {2eV\over h} \left(\frac{L}{\ell_1}\right)^{2/3} e^{-\Delta/T}, 
\end{align}
where $\zeta$ is a numerical coefficient defined through an integral equation 
and numerically found to be $\zeta\approx1.25$,
as discussed in Appendix~\ref{app:ballistic}.
The resulting power-law  dependence $\delta G\propto L^{2/3}$ 
in the wire length $L$ is a consequence of the scaling form of the Kramers
equation and can be understood as follows.~\cite{peq} 

In combination with Eq.~\eqref{bsFP} the power-law scaling
$\dot{N}^R\propto L^{2/3}$ implies that the discontinuity at $p=0$ in
the distribution of right-moving excitations broadens with the
distance $x$ from the left lead as $\partial_p
g_{p,x}|_{p\searrow0}\sim g/\Delta p\sim x^{-1/3}$.  This scaling (and
correspondingly for distribution of left-moving excitations with
distance from the right lead) reflects the diffusive nature of the
backscattering processes.  Excitations entering e.g.\ from the right
lead with momentum $\Delta p$, move to the left, gradually decrease
their velocity in collisions, and eventually return to the right lead.
In order to lose momentum of order $\Delta p$ an excitation has to
experience sufficiently many collisions in the wire, which requires a
time $t$ determined from the standard diffusion condition $(\Delta
p)^2\sim B t$.  Propagating through the wire at a typical velocity
$\Delta p/m^*$ until the turning point, the excitation thus moves a
distance $(\Delta p/m^*)t\sim x$ from the lead.  Combining these two
observations, one obtains $\Delta p\sim (Bm^*x)^{1/3}$, and thus
$\dot{N}^R \propto \int dx (g/\Delta p) \propto L^{2/3}$.

Finally, let us address the crossover from the inhomogeneous diffusive to the quasi-ballistic regime.
The latter is characterized by a length of the wire shorter than the average length scale on which a hole at the bottom of the band 
scatters off low-energy excitations. This regime recently has been studied by Lunde {\it et al.}\cite{Lunde2007} 
for weakly interacting electrons within a perturbative treatment of the Boltzmann equation. This approach builds on the observation that, 
 as the typical highly excited hole participates in at most one collision, 
 backscattering occurs via a single collision and smearing of the 
 hole distribution near $p= 0$ can be neglected. Of course, in this
 regime the picture of diffusive dynamics in momentum space 
 underlying the Fokker Planck approximation does not apply.

To elaborate this point let us recall that the Fokker-Planck approximation relies on a 
gradient expansion of the collision integral.  
The latter applies if the typical momentum 
exchange $q\sim T/v$ in a collision is small compared to the momentum
scale $\Delta p\sim(Bm^*x)^{1/3}$
 on which the distribution 
varies.  Applying this criterion, we 
find that the inhomogeneous diffusive regime 
is limited to length scales larger than $\ell_0 \sim  T^3/(v^3 Bm^*)$. 
Notice that $\ell_0\sim\ell_1 (T/m^*v^2)^{3/2}$ and at low
temperatures $T\ll m^*v^2$
the result \eqref{swl} for the inhomogeneous diffusive regime thus applies within 
a broad region. At the crossover $L\sim \ell_0$
the characteristic scale  $(L/\ell_1)^{2/3}\sim T/(m^*v^2)$, 
and the backscattering rate   
\begin{align}
\label{nrtrans}
\dot{N}^R \sim -{eV\over h}{T\over m^*v^2}e^{-\Delta/ T} 
\end{align}
is thus independent of $B$. 

For weak interactions $\ell_1$ has been calculated from a microscopic
theory.  It was shown to be related to the typical time scale for a
three-particle collision $\tau_{eee}$ as $\ell_1\sim
(\mu/T)^{1/2}v_F\tau_{eee}$.~\cite{peq} Building on this result we
find that at weak interactions the limit of the inhomogeneous
diffusive regime is set by $\ell_0\sim (T/p_F)\tau_{eee}$.  Notice
that $T/p_F$ is the velocity of a hole which can backscatter in a
single three-particle collision with typical momentum exchange $q\sim
T/v_F$. We thus observe that in wires of length $L\lesssim \ell_0$
such hole will typically suffer at most one collision when traversing
the wire, i.e. $L\sim \ell_0$ also defines the limit below which the
quasi-ballistic regime of Lunde {\it et al.}\cite{Lunde2007} applies.
We thus expect that for $L\lesssim \ell_0$ the perturbative
calculation of Lunde {\it et al.} holds.  Indeed, their result
$\dot{N}^R\sim -{eV\over h}{T\over \mu}{L\over \ell_0}e^{-\mu/ T}$ at
$L\sim \ell_0$ matches Eq.~\eqref{nrtrans}.  The linear scaling
$\dot{N}^R\propto L$ here simply reflects the fact that in the
quasi-ballistic regime the probability of backscattering linearly
increases with the time spent in the wire.  As this linear dependence
does not rely on the assumption of weak interactions, one can use
Eq.~(\ref{nrtrans}) to extend the result by Lunde {\it et al.} to
arbitrary interaction strength,
\begin{align}
\label{genLunde}
\dot{N}^R \sim -{eV\over h}{T\over m^*v^2} {L\over \ell_0}e^{-\Delta/ T}.
\end{align}
Precise determination of the numerical prefactor in
Eq.~(\ref{genLunde}) would require a more careful treatment.

\section{Crossover to homogeneous diffusive regime}
\label{sec:longwires}

We next discuss how the result for the inhomogeneous diffusive
regime $\ell_0 \ll L\ll \ell_1$ crosses over into the homogeneous diffusive regime 
$\ell_1\ll L \ll \ell_{\rm eq}$. 
To this end we need to address the full inhomogeneous
Fokker-Planck equation \eqref{fpeq} subject to the boundary conditions
\eqref{bc}.

Following the procedure of the previous section we decompose the
hole-distribution $g_{p,x}$ into a spatially homogeneous and an
inhomogeneous part
\begin{align} 
\label{eq:distr_decomp}
g_{p,x}=g^0_p+\delta g_{p,x},
\end{align} 
where the homogeneous part is readily found as~\cite{peq}
\begin{align}
\label{hs}
g^0_p=e^{\epsilon_p/T}\left[1-\frac{eV}{\sqrt{2\pi m^* T^3}}\int^{p}_{0}dp'\,
e^{-\epsilon_{p'}/T}\right]e^{-\Delta/T}\,.
\end{align} 
The homogeneous distribution \eqref{hs} gives a contribution to the
backscattering rate \eqref{bsFP} that scales linearly with the length
of the wire and dominates at $L \gg \ell_1$.

To find the inhomogeneous solution we return to the dimensionless variables $q$ and $y$ 
in Eq.~\eqref{dimlessv} and start out from the ansatz 
\begin{align}
\delta g_{q,y}
&=
e^{a (y-q)}h(q).
\end{align}
This leads us to the  differential equation for $h$ 
\begin{align}
\label{dh}
\partial^2_q h
-
2(q+a)\partial_q h
+
2\left(-1+\frac{a^2}{2}\right)h=0,
\end{align}
which for the special values of the parameter,
 $a_n=\pm \sqrt{2(n+1)}$ and $n=0,1,... $  
 is solved by Hermite polynomials with shifted arguments $H_n(q\pm a_n)$. 
Using again that in the linear response regime $\delta g_{q,y=0}$ is antisymmetric in $q$
and noting that $H_n(q)=(-1)^nH_n(-q)$,  
the general solution to \eqref{fpeq} reads 
\begin{align} 
\label{inh}
\delta g_{q,y} 
 &=
 \sum_{n=0}^{\infty} b_{n} 
 \left( e^{a_n y } \psi_n(q) - e^{-a_n y} \psi_n(-q)
\right), 
\end{align}
where
\begin{align} 
\label{ef}
 \psi_n(q) 
 &= \frac{1}{\sqrt{\mathcal{N}_n}} e^{-a_n q} H_n (q+ a_n),
\end{align}
and we introduced the normalization constant 
$\mathcal{N}_n=\sqrt{\pi} a_n 2^n n! e^{2(n+1)} $. 
 Expansion coefficients $b_n$ are found 
from matching ansatz \eqref{inh} to the boundary conditions, 
\begin{align}
\label{bc3}
 \theta (q) g_{q,-\Lambda} + \theta (-q) g_{q,\Lambda} 
 &= 
 -
{eV\over 2T} e^{q^2-\Delta/T}(\sgn(q)-\erf(q)), 
\end{align}  
and details of this calculation can be found in Appendix~\ref{app:crossover}.
Accounting then for homogeneous and inhomogeneous contributions 
to the distribution function
the backscattering rate reads 
\begin{align}
\label{eq:backsc}
\dot N^{R}= - {2eV\over h} \left( \frac{L}{\sqrt{\pi}\ell_1} 
+  I^t \left( \openone + OX \right)^{-1} X  I \right) 
 e^{-\Delta/T} \, ,
\end{align}
where we introduced vector $I$ and matrices $X$, $O$, respectively, 
with coefficients
\begin{align}\label{eq:coeff}
 I_m 
 &= -  \psi'_m(0)/a_m,  
 \\
 X_{mn} 
 &= \delta_{mn}  \left(1 - e^{-2a_n \Lambda} \right)/2, 
 \\
 O_{mn}
 &=
 \label{eq:c}
 \begin{cases}   
 c_{mm} + \int_0^{\infty} dp^2 \,  e^{-p^2-a_m^2}\psi^2_m(p), 
 \quad 
 m=n
 \\
 {2  \over a^2_m  - a^2_n }
\left( 
a_m^2 c_{mn} - a_n^2 c_{nm}
\right),  
\quad  
m\neq n
 \end{cases}
\end{align}
and $c_{mn}=I_m \psi_n(0)$.

In the inhomogeneous diffusive regime $\ell_0\ll L\ll\ell_1$ main contributions to 
Eq.~\eqref{eq:backsc} originate from coefficients with large index $n\gg \ell_1/L$.
This allows to approximate Hermite polynomials in the eigenfunctions \eqref{ef}  
by Airy functions according to~\cite{Dominici2006}  
$H_n(z)\simeq \sqrt{2\pi(2n)^n}n^{1/6}\mathrm{Ai}\left(\sqrt{2}n^{1/6}(x-\sqrt{2n})\right)
e^{-\frac{3}{2}n+\sqrt{2n}z}$. 
Reassuringly, upon this substitution one recovers Eq.~\eqref{swl}.

In the opposite limit $L \gg \ell_1 $ one may approximate $2X\simeq \openone$ and thus
finds the backscattering rate 
\begin{align} 
\label{nrgen}
 \dot N^R = - {2eV\over h} \left( 
 {L\over \sqrt{\pi}\ell_1}
+ \xi 
\right) e^{-\Delta/T},
\end{align}
with an universal offset numerically calculated as $\xi \simeq 0.275$.

\begin{figure}[tb]
\centering
\resizebox{.47\textwidth}{!}{\includegraphics{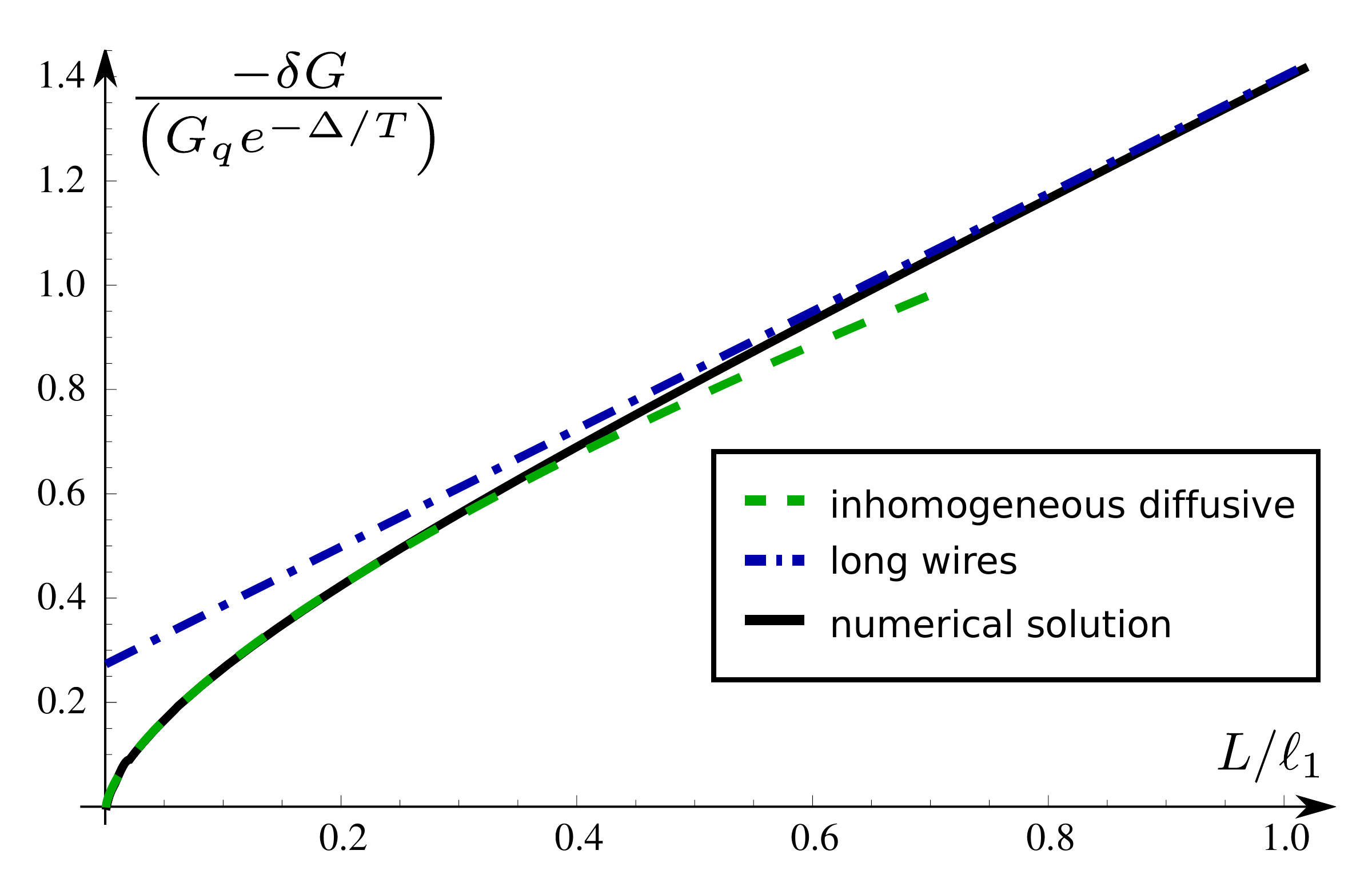}}
\caption{
Length dependence of the interaction-induced correction to the conductance Eq.~\eqref{gnon},
found from the backscattering rate of quantum wires $\ell_0\ll L\ll \ell_{\rm eq}$.  
Solid line shows the result obtained from numerically solving Eqs.~\eqref{eq:backsc}-\eqref{eq:c}. 
Asymptotic results for wires much shorter and much longer than $\ell_1$,
 Eqs.~\eqref{swl} and~\eqref{nrgen} respectively, are indicated by dashed and dash-dotted lines. 
 The characteristic scaling $\delta G\propto L^{2/3}$ in the 
 inhomogeneous diffusive regime holds up to $L\lesssim 0.4 \ell_1$,  
 and then crosses over to the result for the homogeneous diffusive regime 
 with its linear length dependence and universal off-set $\xi\simeq 0.275$. 
}
\label{fig4}
\end{figure}

To address the crossover regime at arbitrary ratios
$L/\ell_1$ we numerically evaluated \eqref{eq:backsc}.  The resulting
backscattering rate leads to a finite temperature correction to the conductance \eqref{gnon}
 shown in Fig.~\ref{fig4}.  We observe that the characteristic power-law $\dot{N}^R\propto L^{2/3}$ 
 of the inhomogeneous diffusive regime extends up to  lengths of the wire $L\lesssim 0.4 \ell_1$. 
 Once the wire length exceeds $L\gtrsim 0.8 \ell_1$, the correction follows the linear length dependence of the homogeneous diffusive regime 
 with the universal off-set $\xi \simeq 0.275$.  
 These features hold for weak as well as strong interaction, and the peculiar power law 
 dependence of $\delta G$ on the wire length is, 
therefore, characteristic to short clean quantum wires $\ell_0\ll L\ll \ell_{\rm eq}$.

\section{Summary and discussion}

We have studied the interaction-induced backscattering rate
$\dot{N}^R$ in a voltage-biased clean quantum wire, in which
conservation laws suppress relaxation.  Our calculations 
apply for wires in
a broad range of lengths $\ell_0\ll L\ll \ell_{\rm eq}\propto
e^{\Delta/T}$, which are short in the sense that equilibration has not
fully established, but long enough to guarantee diffusive motion in
momentum space due to interaction-induced collisions.  In these wires,
a finite rate $\dot{N}^R$ arises due to the backscattering of mobile
holes at the band bottom performing the random motion of a Brownian
particle in momentum space while scattering from excitations at the
Fermi level.  Reminiscent to the first passage problem the dynamics of
the hole is described by an inhomogeneous Fokker-Planck equation.

From solutions of the latter we have derived the wire length
dependence of the interaction-induced correction to the conductance
\eqref{gnon}, and found a power law scaling $\delta G \propto L^{2/3}$
as a characteristic feature of these wires.  We have identified the
length scale $\ell_0$ which separates the diffusive from the
quasi-ballistic regime. The latter has previously been studied by
Lunde {\it et al.}~\cite{Lunde2007} in the limit of weak interactions.
We confirmed that for weakly interacting electrons backscattering
rates in both regimes match at the crossover scale $L\sim\ell_0$, and
were able to generalize the results by Lunde {\it et al.}  to
arbitrary interaction strengths.

Our results hold for weakly as well as strongly interacting electrons.
They depend on the interaction strength via the bandwidth of
excitations $\Delta$, which sets the activation energy
$e^{-\Delta/T}$, and effective mass $m^*$ and diffusion constant $B$,
which both define the relevant length scale of the problem,
$\ell_1$. The activation behavior $e^{-\Delta/T}$ dominates the
temperature dependence of the backscattering rate and corresponding
correction to the conductance.  An additional temperature dependence
enters via the diffusion constant defining the pre-exponential factor.
The diffusion constant has been studied for spin-polarized electrons
at arbitrary interactions~\cite{peq,eqWC,eqLL,Matveev2012} where a
temperature dependence $B\sim T^5$ was found.  While a generalization
accounting for spin degree of freedom and applicable at arbitrary
interaction strength is still an open problem, the strongly
interacting limit of a Wigner crystal has been addressed
recently.~\cite{MatveevAndreevKlironomos} There, a scaling $B\sim T^3$
was found, and a similar result holds at arbitrary
interaction strength.~\cite{to be published} 

\subsection*{ Acknowledgments} 
K.~A.~M. and T.~M. acknowledge fruitful discussions with J.~Rech. Work
by M.~T.~R. was supported by the Alexander von Humboldt Foundation.
T.~M. acknowledges support by Brazilian agencies CNPq and FAPERJ.
Work by K.~A.~M.  was supported by the U.S. Department of Energy,
Office of Science, Materials Sciences and Engineering Division. 
A.~L. acknowledges support from NSF grant DMR-1401908 and GIF, and thanks 
I.~V.~Gornyi and D.~G.~Polyakov for numerous important discussions. 

\begin{appendix}

\section{Orthogonality relations} 
\label{app:fokpla}

The inhomogeneous solution to the FPE \eqref{fpeq} is expanded in functions 
$\psi_n( \pm q)$ which are eigenfunctions of the differential operator
\begin{align} 
\label{appd}
\mathcal{D} ={1\over q} \partial_q \left( {1\over 2} \partial_q - q\right),
 \end{align}
i.e. $\mathcal{D} \psi_m(sq) =  sa_m \psi_m (sq)$, with $s =\pm$, 
and are orthogonal with respect to a 
weight function $ w(q) $ to be determined in the following way. 
From the eigenvalue problem it follows that
\begin{align}
 &\psi_n(s q) 
  \mathcal{D} 
  \psi_m (s' q) - \psi_m(s' q) 
  \mathcal{D} \psi_n (s q) 
 \nonumber \\
 &\qquad 
 = \left(s' a_m - s a_n \right)\psi_n (s q) \psi_m (s' q),
 \quad s,s'=\pm.
\end{align}
Multiplying both sides with a, yet to be determined, function $w(q)$, integrating over the 
entire momentum range and imposing an orthogonality condition 
\begin{align}
\label{orthog}
\int_{-\infty}^\infty dq\, 
w(q)\psi_n(s q)\psi_m( s' q)
&= 
- s 
\delta_{ss'}\delta_{n,m},
\end{align}
we find the weight function to be determined by the differential equation 
\begin{equation}
-{1\over 2q}\partial_qw(q)
+
w(q)\left( 
{1\over 2q^2}-1
\right)=0,
\end{equation}
resulting in $w(q)=qe^{-q^2}$. 

In the inhomogeneous diffusive regime the differential operator of interest reads
\begin{align}
\mathcal{D} 
&=
{1\over 2q}\partial^2_q,
\end{align}
and following the same steps as above one arrives at 
the orthogonality relation for Airy-functions
\begin{align}
\label{orthAi}
 \int_{-\infty}^\infty dq \, q {\rm Ai}(\alpha q) {\rm Ai}(\beta q) 
= -{1\over 3\alpha} \delta(\alpha-\beta).
\end{align}

\section{Details on the inhomogeneous diffusive regime} 
\label{app:ballistic}

We give details on the derivation 
of the expansion coefficients defining 
the hole distribution function in \eqref{anskeq}
and the backscattering rate in the inhomogeneous diffusive regime.

Coefficients $b(a)$ are found from matching 
\eqref{gairy} to the boundary condition 
\begin{align}
 \theta (q) g_{q,-\Lambda} + \theta (-q) g_{q,\Lambda} 
 &= -e^{-\Delta/T} \frac{eV}{2T} \sgn (q),
\end{align}
Inserting the explicit form of the general solution $g_{q,y} $
and expressing $e^x=\cosh x+\sinh x$ one finds
\begin{align}
\label{bcexp}
 \int_{0}^\infty 
 &da
 { b(a)\over (2a)^{2/3}}  
 \left[ 
 e^{a \Lambda} \AiryAi ( (2a)^{1/3} q)
 -  e^{-a \Lambda} \AiryAi (- (2a)^{1/3} q) \right.
 \nonumber \\ 
& \left.
-
2\Theta(q) \sinh(a\Lambda)
\left( \AiryAi ( (2a)^{1/3} q)  + \AiryAi (- (2a)^{1/3} q) \right)
\right]
 \nonumber \\ 
&
= -e^{-\Delta/T} \frac{eV}{2T} \sgn(q). 
\end{align}

Coefficients $b(a)$ can then be extracted
using the orthogonality relation for Airy-functions \eqref{orthAi},
i.e. by multiplying left and right hand side
of \eqref{bcexp} with `$q \AiryAi\left((2a_0)^{1/3}q\right)$' where $a_0>0$ and integrating over all $q$.
For the right hand side of  \eqref{bcexp} we may further 
use that
\begin{align}
 \int_{-\infty}^\infty dq\, 
|q|\AiryAi\left((2a_0)^{1/3}q\right)
&= 
- 2^{1/3} a_0^{-2/3} \AiryAi'(0) 
\end{align}
where we used the defining differential equation for Airy functions, to 
e.g. 
calculate 
$\alpha^3 \int_0^\infty dq\, q\AiryAi(\alpha q)
=
\int_0^\infty dq\,\partial_q^2\AiryAi(\alpha q)
=-\alpha \AiryAi'(0)$.
We then find  
\begin{widetext}
 \begin{align}
 {b(a_0) e^{a_0\Lambda}\over (16 a_0)^{1/3}}
 +
 \int_0^\infty da \, 
 {b(a)\over (2a)^{2/3}}\sinh(a\Lambda) {\cal F}(a,a_0)
 &=
 -{2^{1/3} \AiryAi'(0) \over a_0^{2/3}}  {eV\over 2T}e^{-\Delta/T},
 \end{align} where
 \begin{align}
\label{opro}
 {\cal F}(a,a_0)
 &=
 2\int_0^\infty dq \, 
 q
 \AiryAi ( (2a_0)^{1/3} q) 
\left(
  \AiryAi\left((2a)^{1/3}q\right)
+  \AiryAi\left(-(2a)^{1/3}q\right)
\right).
\end{align}
\end{widetext}
The function ${\cal F}$ can be further 
calculated with help of the identity
\begin{align}
&\left[ 2a- s 2a_0\right]  
\int_0^\infty dq\, 
q \AiryAi\left((2a)^{1/3}q\right) 
\AiryAi\left(s(2a_0)^{1/3}q\right)
\nonumber\\
&\qquad 
=
-\left[ (2a)^{1/3}- s (2a_0)^{1/3}\right] \AiryAi'(0)\AiryAi(0),
\end{align}
which again results from using the defining differential equation 
for the Airy function to express 
$q\AiryAi(\alpha q)= \alpha^{-3} \partial_q^2\AiryAi(\alpha q)$, 
and gives
\begin{align}
 {\cal F}(a,a_0)
 &=
- \AiryAi'(0)\AiryAi(0)
{ (2a)^{4/3}- (2a_0)^{4/3}\over a^2 - a_0^2}.
\end{align}

We thus find
 from orthogonal projection
the integral equation 
\begin{align} 
 \label{app_eq:1}
 \beta(a_0\Lambda) 
 +
{a_0^{2/3}\over \sqrt{3}\pi} 
\int_0^\infty {da \over a} \beta(a \Lambda) \left(1 - e^{- a \Lambda} \right) 
{a_0^{4/3}-a^{4/3}\over a_0^2-a^2}
 =
  1,
 \end{align} 
where we introduced 
\begin{align}
\label{bbeta}
 b(a)
 &= -{ 2^{2/3} {\rm Ai}'(0) \over a^{1/3}} 
 {eV \over T}   
e^{-\Delta/ T} e^{-a \Lambda}
 \beta(2a\Lambda),
\end{align}
and employed that 
$\AiryAi'(0)\AiryAi(0)=-\left(2\sqrt{3}\pi\right)^{-1}$.
Rescaling $a$ and $a_0$ by the factor $1/\Lambda$ one arrives at the 
expression stated in the main text.

Finally, the backscattering rate expressed in dimensionless variables
\begin{align}
\dot{N}^R 
&= 
{2T\over h}
\int_{-\Lambda}^\Lambda dy\, 
 \left( \partial_q g_{q,y} \right)|_{q=0}
\end{align}
is readily calculated as  
\begin{align}
\dot{N}^R 
&= 
2^{8/3} 
\AiryAi'(0) 
{ T\over h} 
\int_0^\infty {da\over a^{4/3}} b(a) 
\sinh(a\Lambda),  
\end{align}
or expressed in terms of $\beta(a)$ 
(see Eq.~\eqref{bbeta})
\begin{align}
\dot{N}^R 
&= 
-
\left( 2\AiryAi'(0) \right)^2
{2eV \over h}
e^{-\Delta/T}
\int_0^\infty {da \over a^{5/3}} \beta(a\Lambda) 
\left(1- e^{-a\Lambda} \right).   
\end{align} We may now scale $\Lambda$ out of the integral 
and find the result Eq.~\eqref{swl} stated in the main text, 
with a 
numerical constant 
$\zeta 
= {(4/3)^{2/3} \over \Gamma^2(1/3)} 
\int_0^\infty da {\beta(a)\over a^{5/3}} \left(1- e^{-a} \right)$, 
where the latter involves a solution of the integral equation,
\begin{align}  
\label{appbet2}
&\beta(a_0)
+ \gamma a_0^{2/ 3}   
  \int_0^\infty {da \over a} \left( 1- e^{-a}\right) {a^{4/ 3}-a_0^{4/ 3}\over
a^2- a_0^2}  \beta(a)
 =1,
\end{align}
and $\gamma = 1/\sqrt{3}\pi$. 
Solving \eqref{appbet2} numerically we find $ \zeta~\simeq~1.25$.

\section{Details on the crossover regime}
\label{app:crossover}

We give details on the derivation 
of the expansion coefficients defining  
the hole distribution function and the backscattering rate in the crossover regime.

Coefficients $b_n$ entering the general solution \eqref{inh} are derived from the 
boundary conditions  discussed in the main text in a similar way as discussed in the previous section.
Employing the orthogonality relation \eqref{orthog} for the general eigenfunctions and 
proceeding analogously as in the inhomogeneous diffusive regime
we arrive at the following 
equation, generalizing \eqref{app_eq:1} 
to the crossover regime,
\begin{align} 
\label{lgs}
   \beta_n
   + {1\over2}  \sum_{m=0}^\infty O_{nm}  
   \left( 1- e^{-2 a_n \Lambda} \right) \beta_m 
&=  
I_n.
\end{align}
Here we introduced (similar to \eqref{bbeta})
\begin{align}
\label{apbn}
b_n =  {eV\over 2T} e^{ - \Delta/T} e^{- a_n \Lambda} \beta_n,
\end{align}
and vector and matrix elements $I_m$  and $O_{mn}$, respectively,
are defined as
\begin{align}
\label{eq:coefficients1}
I_n 
&=  
\int_{-\infty}^\infty dq \, 
q \psi_m(q) \left(\sgn(q)- \erf(q)  \right), 
\\
O_{mn} 
 &= 
 2 \int_0^\infty dq \, 
 q  e^{-q^2} \psi_m(q)   
 \left(\psi_n(q) + \psi_n(-q) \right). 
\end{align}

The above integrals can 
be evaluated using eigenvalue equations below \eqref{appd}. 
We may express vector elements e.g. as
\begin{align}
 I_m 
 &= 
 {1\over a_m} 
 \int_{-\infty}^\infty dq \,  
 \mathcal{L} \psi_m(sq)  
 \left(\sgn(q)- \erf(q)  \right),
\end{align}
with $\mathcal{L} = q \mathcal{D}$ and ${\cal D}$ from Eq.~\eqref{appd}, 
and upon integration by parts (notice that boundary terms vanish) find 
\begin{align} 
\label{eq:coeff_In_2}
 I_m 
 = -{1\over a_m} 
 &\Big[
 \left( 
  \partial_q \psi_m(q) -2q \psi_m(q) 
  \right)|_{q=0}  \nonumber \\
 & 
  -\int_{-\infty}^\infty {dq\over \sqrt{\pi}} 
  \partial_q \left(e^{-q^2} \psi_m(q)\right)  
 \Big],
\end{align}
as stated in the main text. 

In a similar way we calculate integrals defining matrix elements $O_{mn}$
starting out from
\begin{align}
\int_0^\infty dq \, 
q e^{-q^2} \psi_m(q) \psi_n(sq) 
&= 
  {1\over a_m} \int_0^\infty dq \, 
  e^{-q^2} \mathcal{L} \psi_m(q) \psi_n(sq). 
\end{align}
Upon partial integration and further algebraic manipulations we arrive
at
\begin{align}
\label{appis}
& 2\int_0^\infty dq \, 
 q  e^{-q^2} \psi_m(q) \psi_n(sq) 
 \nonumber\\
 &\qquad = 
{1\over a_m - s  a_n}
 \left(  
 s \psi_m(0) \psi'_n(0)
- 
\psi_n(0) \psi'_m(0) 
\right), 
\end{align}
which applies for all $m$, $n$ ($m \neq n$) if $s=-$ ($s=+$), and 
leads to the result stated in the main text.

Finally, the backscattering rate is found from the general expression,   
 \begin{align}
 \dot{N}^R
 &={8T\over h}\sum_n {b_n\over a_n}
 \psi_n'(0) \sinh(a_n\Lambda),
 \end{align}
upon introducing 
$I_n=\psi'(0)/a_n$
 and inserting the formal solution
to Eq.~\eqref{lgs}, 
 \begin{align} 
\label{eq:coeff_matrix}
 \beta_n 
 &= 
\sum_m \left( (1+OX)^{-1} \right)_{nm} I_m, 
 \end{align}
with $\beta_n$ defined in Eq.~\eqref{apbn}.

\end{appendix}


\begin{thebibliography}{99}
\bibitem{remark2}
Nonequilibrium corrections to Fokker-Planck equation \eqref{fpeq} affecting states near the Fermi points 
have been derived and analyzed in the context of Coulomb drag problem, 
see Ref.~\onlinecite{DGP} for extensive details.
\end{thebibliography}
\end{document}